\documentstyle[epsfig]{mn}
\newcommand{\be}{\begin{equation}}
\newcommand{\ee}{\end{equation}}
\newcommand{\ba}{\begin{eqnarray}}
\newcommand{\ea}{\end{eqnarray}}

\newcommand{\err}{\end{array}}
\newcommand{\bc}{\begin{center}}
\newcommand{\ec}{\end{center}}

\newcommand{\hm}{\,h^{-1}{\rm Mpc}}

\newcommand{\etal}{et al.~}
\newcommand{\eg}{e.g.,~}
\newcommand{\ie}{i.e.,~}

\title[Shapes and Sizes of Voids]{Shapes and Sizes of Voids in the
  $\Lambda$CDM Universe: Excursion Set Approach } 
\author[Shandarin \etal]{Sergei Shandarin$^\dagger$$^\ddagger$, Hume
  A. Feldman$^{\dagger}$, Katrin Heitmann$^\star$ \& Salman Habib$^\ddagger$\\
  $^\dagger$Department of Physics \& Astronomy, University of Kansas,
  Lawrence, KS 66045, USA\\
  $^\star$ISR-1, The University of California, Los Alamos National
  Laboratory, Los Alamos, New Mexico 87545, USA\\
  $^\ddagger$T-8, The University of California, Los Alamos National
  Laboratory, Los Alamos, New Mexico 87545, USA\\
  emails: sergei@ku.edu;\, feldman@ku.edu;\, heitmann@lanl.gov;
  \,habib@lanl.gov}
\begin{document}
{}
\date{\today, LA-UR-05-4302}

\maketitle

\begin{abstract}
  We study the global distribution and morphology of dark matter voids
  in a $\Lambda$CDM universe using density fields generated by N-body
  simulations. Voids are defined as isolated regions of the
  low-density excursion set specified via density thresholds, the
  density thresholds being quantified by the corresponding filling
  factors, i.e., the fraction of the total volume in the
  excursion set. Our work encompasses a systematic investigation of
  the void volume function, the volume fraction in voids, and the
  fitting of voids to corresponding ellipsoids and spheres. We
  emphasize the relevance of the percolation threshold to the void
  volume statistics of the density field both in the high redshift,
  Gaussian random field regime, as well as in the present epoch. By
  using measures such as the Inverse Porosity, we characterize the
  quality of ellipsoidal fits to voids, finding that such fits are a
  poor representation of the larger voids that dominate the volume of
  the void excursion set.
\end{abstract}

\begin{keywords}
{voids, dark matter,  N-body, excursion set}
\end{keywords}

\section{Introduction}
\label{sec:intro}

The detailed distribution of mass in the Universe is only now being
fully revealed by large-scale astronomical surveys. Consequently, in
the last decade or so, a great deal of effort has gone into detailed
studies of the high density regions, in particular the progenitors of
galaxies, and clusters of galaxies. The density profiles, merging
history, substructure and many other features have been increasingly
accurately modeled, simulated, and analyzed.  This is not surprising
because these are precisely the places where most ``action'' occurs:
the first objects, galaxy formation and merging, X-rays, gravitational
lensing, etc.

But even the lower-density, ostensibly ``quiet'' regions of the
Universe are of significant interest: recent studies indicate a
discrepancy between observations and theory in low and medium density
regions which are naturally associated with voids and superclusters
respectively. Simulations show that dark matter halos in voids are
capable of developing into void objects observable as (dwarf)
galaxies.  However, the morphology-density correlation so natural in
the biased galaxy formation scenario seems to be in contrast to what
is observed.  Despite a fairly long history of studies [see, \eg
Zel'dovich, Einasto, \& Shandarin (1982), Rood (1988)] the problems of
void phenomenology remain serious enough for Peebles (2001) to
conclude: ``The apparent inconsistency between the theory and
observations of voids is striking enough to be classified as a crisis
of the CDM model.''

The study of low and medium density regions is demanding from both the
observational and modeling perspectives. From the point of view of
observations, large contiguous regions need to be surveyed in
sufficient depth and spectra, colors, and morphology information for
the surveyed galaxies obtained. Fortunately, the observational
impediment is coming to an end: the 2dF redshift survey~\cite{2df} has
been completed and the data released, and the Sloan Digital Sky Survey
(SDSS) has begun to release large volumes of data~\cite{sdss}.  As a
result, valuable information on the luminosity function, colors,
concentration and other attributes of galaxies in voids and void walls
have been obtained \cite{roj_etal_03,hoy_etal_03}. The interpretation
of these results, however, relied heavily on the purely
phenomenological VOIDFINDER algorithm of Benson~\etal (2003) (see also
references therein) that assumes voids to be nearly spherical and also
depends on ad hoc parameters such as the criteria for ``void'' and
``wall'' galaxies first introduced by El-Ad \& Piran (1997).  As a
cautionary remark, the relation of the above observational results to
the physical processes in voids and void walls does not seem to be
trivial.

Observational studies of voids and superclusters searching for various
correlations between properties of galaxies and their environment
often do not take into account the fact that some galaxies may travel
distances comparable to the sizes of voids (up to roughly half of the
radius of a spherical void, as estimated in Appendix 1), and are then
artificially shifted further by the observationally unavoidable
mapping into redshift space.  Therefore the environment where the
galaxies were originally formed might well be quite different from the
one where they are ostensibly observed: The physical processes that
govern the observed properties of galaxies in voids and in void walls
certainly operate in physical space and not in redshift space.

Additionally, theoretical models of voids remain oversimplified; the
assumption that voids have close to spherical shapes is often the
basis of theoretical and observational
studies~\cite{kau_fai_91,dub-etal-93,ryd_95,ame_etal_99,gol_vog_04}.
Some algorithms formally allow nonsphericity but in a quite limited
form (see \eg Aikio \& M\"{a}h\"{o}nen 1998).  This prejudice is
partly based on a result due to Icke (1984) who noticed that by
changing the sign in the solution for the collapse of a uniform
ellipsoidal enhancement (Lin, Mestel \& Shu 1965) one can describe the
evolution of an ellipsoidal void. He showed that the asphericities of
an isolated ellipsoidal underdense perturbation ``tend to disappear''
as the void becomes bigger. Following this, many theoretical studies
of void evolution have taken the spherical model for granted. Icke's
conclusion, however, was based only on inequalities regarding the
relative accelerations along principal axes of the ellipsoid. He did
not integrate the equations and therefore the evolution of the ratios
of the principal axes remained unknown.  Bertschinger (1985), who
studied the evolution of an isolated uniform ellipsoidal void
numerically, confirmed the effect found by Icke.  However,
quantitatively his result was not very impressive: an initially oblate
ellipsoid with axis ratio $c_i/a_i=0.5$ and density
$\rho_i=0.9\bar{\rho}(t_i)$ ($\delta_i=-0.1$) evolved to a more
spherical ellipsoid with $c_f/a_f=0.722$ when the density inside the
ellipsoid dropped to $ \rho_f=0.04\bar{\rho}(t_f)$ ($\delta_f=-0.96$)
after the Einstein-de Sitter universe had expanded to 162 times its
initial size. One can hardly expect that a void can evolve for a such
long time without being seriously disturbed by its neighbors. For
instance, as predicted by the adhesion approximation, the shape of
voids is likely to be affected by violent interactions with adjoining
superclusters and voids resulting in the growth of some voids at the
expense of other voids \cite{kof_etal_92,sah_sat_sh_94}. Another
caveat regarding Icke's model is the negligence of external shear.
Eisenstein \& Loeb (1995) have emphasized that even in the case of the
collapsing ellipsoid the shape change is primarily induced by the
external shear and not by the initial triaxiality of the objects. Due
to their low internal density, it is clear that the external shear
must play an even stronger role in the case of voids.  In any case,
the issue of dynamical evolution of voids appears to be unsettled and
requires further study.

From the point of view of detailed modeling, voids require large
simulation volumes, while at the same time, high mass resolution is
necessary to track the smaller dark matter halos that populate the
voids. In addition, hydrodynamics, feedback, star formation, etc. need
to be modeled adequately to understand galaxy formation in the void
complex. It is worth noting here that
theory~\cite{kof_etal_92,sah_sat_sh_94} as well as some N-body
simulations~\cite{kof_etal_92,wey_kam_93,dub-etal-93,got_etal_03}
indicate that both superclusters and voids have nontrivial
substructure and complicated internal dynamics.  As an example, very
complex substructures have been detected and studied in dark matter
superclusters~\cite{sh_she_sah_04} and voids by the Virgo consortium
utilizing N-body simulations~\cite{jen_etal_98}.

The problems outlined above require a more systematic study of voids
-- of their morphology and topology, dynamics, and internal structure.
Assuming that the low dark matter density regions are closely related
to observational voids, we concentrate here on the most basic global
properties: their sizes and shapes.  We study the dark matter mass
distribution in real space obtained from N-body simulations of the
$\Lambda$CDM model.  Although the mass distribution in real space
cannot be measured directly, it controls the dynamics, as well as
other properties of voids.

It is worth stressing that there is no commonly accepted definition of
a void. Here we employ the underdense excursion set approach: voids
are defined as individual regions where the density contrast is below
a certain threshold $\delta_c$. The boundaries of the voids are
therefore closed surfaces of constant density contrast $\delta_c$. The
advantages of this definition include: (i) simplicity, (ii) a clear uniform
definition of void boundaries 
\footnote{Numerous void finding algorithms  often define different parts of the boundary 
surface by different conditions.} 
allowing the study of shapes and sizes,
and (iii) symmetry with the common definition of superclusters as
regions with densities above a specified density threshold.  We do not
preselect any particular threshold but rather study the transformation
of voids as the threshold is raised from low to high values. As in
other definitions of voids, we also assume a certain smoothing
procedure for the density field which we take here to be a uniform
Gaussian filter. This allows a continuous transition to a Gaussian
density field if the scale of the filter window is taken to be
sufficiently large.

As further discussed below, one of the consequences of this definition
is the associated percolation transition: at thresholds above a
certain value the largest void spans the entire volume and quickly
becomes the dominant structure in the underdense excursion set. At the
threshold corresponding approximately to the percolation transition,
the individual voids reach their largest sizes and volumes.  We will
quantitatively study the shapes and sizes of underdense regions at
these thresholds.

Although we do not attempt to review the entire body of work on voids,
we would like to emphasize that many other definitions of voids have
been previously suggested~\cite{kir_etal_81,kau_fai_91,sah_sat_sh_94,ela_pir_97,aik_mah_98,pli_bas_02,ben_etal_03,col-etal-05}. All of them
are claimed to be in agreement with visual impressions of observed
data and N-body simulations. While ``visual impression'' pertains more
to the expertise of psychology rather than physics or astronomy, in
general these various definitions agree at least on one basic
principle: dynamically, voids are closely related to the underdense
regions.  Often the algorithms finding voids are designed to suppress
narrow tunnels between two fat voids, see \eg Kauffmann \& Fairall
(1991); Colberg et al. (2005). This inevitably involves ad hoc
parameters describing which tunnel is narrow and which is not.  In
addition, there are attempts to incorporate some concepts of
theoretical dynamical models into the construction of voids, see, \eg
Dubinski et al. (1993); van de Weygaert \& van Kampen (1993); Benson
et al. (2003); Colberg et al. (2005); in practice this is essentially
the spherical model of void evolution.

In this paper, we develop an unbiased (\ie without any theoretical or
aesthetic or other ad hoc assumptions) technique that is able to
identify voids and then quantify -- at least crudely -- their
shapes. The only `arbitrary' parameter is the scale of smoothing of
the dark matter density field.  The smoothing filter is taken to be a
uniform Gaussian window [$\propto \exp(-r^2/2R_f^2)$].  Obviously,
this is not an optimal smoothing method and we are planning to improve
this part of the technique in future work.  A particularly promising
method of constructing the density field from point-like distributions
obtained in redshift surveys and N-body simulations is based on
Delaunay tesselations~\cite{sch-wey-00,wey-02}.  Density fields
reconstructed in this manner appear to preserve anisotropic features
and therefore may have some advantages over conventional
`Cloud-In-Cell' techniques followed by an isotropic smoothing, as
performed here.

As stated earlier, we define voids as isolated regions of the low
density excursion set specified by a density threshold: $\delta \equiv
\delta \rho/\rho < \delta_c$.  This definition is significantly
different from the many definitions used in the literature in one
particular sense. It defines the boundary of the void explicitly and
unambiguously. In contrast, most other studies define a void as an
underdense region surrounding the minimum of the density field (\eg
van de Weygaert \& van Kampen 1993).  The boundary may remain
unspecified exactly (\eg Gottl\"ober et al. 2003; Sheth \& van de
Weygaert 2004) or be constructed (\eg Kauffmann \& Fairall 1991; Aikio
\& M\"ah\"onen 1998). Though in both cases it is implied that the
boundary approximately corresponds to the surface of constant density,
various exceptions from the rule are usually allowed.

In our systematic study of the void excursion sets at a large number
of density thresholds, we identify a particular threshold when the
voids are the largest by volume. This threshold is just below the
percolation threshold ($\delta \le \delta_{c}$) where voids merge with
each other and form a single percolating void that spans the entire
volume of the simulation~\cite{she_etal_03,sh_she_sah_04}. The
percolating void itself does not have a well-defined volume and has a
very complicated geometry and topology, however, it possesses
interesting scaling properties \cite{bak-med-sh-05}.
We show that the individual regions of the underdense excursion set
do not have spherical shapes: the mean ratio of the smallest to largest
axis is about 0.45 more or less independently of the void 
volume.
In addition, using quantitative measures such as the Inverse Porosity,
we investigate to what extent ellipsoidal and spherical volumes can be
used as good approximations for realistic void shapes (as given by our
definition). We conclude that while such an approach may work for
small voids, it is a poor approximation for the larger voids which
dominate the volume of the void excursion set.

The paper is organized as follows: In \S \ref{sec:nbody} we describe
the N-body simulations used in this study. In \S \ref{sec:VolFunc} we
define the void volume function, the number of voids of a given
volume, and the void volume fraction.  In \S \ref{sec:gauss} we
briefly discuss voids in a cosmologically relevant Gaussian random
field and then, in \S \ref{sec:n-body}, the voids resulting from
N-body simulations of the $\Lambda$CDM model. In \S
\ref{sec:fitellipsoid} we show how to approximate voids by ellipsoids
and define the voids' Inverse Porosity. In \S \ref{sec:sphericity} we
define and quantify the sphericity of voids and discuss the
statistical distribution of void shapes. In \S \ref{sec:shape} we
discuss the shapes of the voids in the simulations. Specifically, we
discuss the mean Sphericity and Inverse Porosity (\S
\ref{subsec:mean}) and their distribution functions (\S
\ref{subsec:distfun}) as found from the simulations. We summarize and
conclude in \S \ref{sec:discussion}. 
 
\section{N-body simulations}
\label{sec:nbody}

The dark matter density field at early times is taken to be a Gaussian
random field, specified completely in real space by its two-point
statistics and in $k$-space, by the primordial fluctuation power
spectrum $P(k)$, typically assumed to be a power-law. At later times,
the density field on the length scales of interest here is nonlinear
and non-Gaussian.  Accurate evolution of the initial density field on
these scales requires the application of numerical N-body methods.

Our primary interest here is in the shapes and sizes of void regions
and not in studying the detailed dynamics and galaxy formation within.
Since the relevant void sizes are on the scales of tens of Mpc, a
simulation box of 256~$h^{-1}$Mpc ($h$ is the Hubble parameter in
units of 100~km/s/Mpc) provides an adequate total volume for the
gathering of void statistics. A void is defined operationally by
considering a smoothed density field with a smoothing scale of the
order of 1~Mpc; it follows that for the studies reported here, force
resolution demands are modest and easily met. Similarly, as we are not
interested in the details of the density field interior to the void,
the mass resolution required is also easily attained. N-body
simulations with N$_p=256^3$ particles corresponding to an individual
particle mass of 1.227$\cdot 10^{11}$~M$_\odot$ are sufficient for the
task.

Given the requirements above, a particle-mesh (PM) code is completely
adequate. The simulations were performed using the parallel PM
code-suite MC$^2$ ({\bf M}esh-based {\bf C}osmology {\bf C}ode), one
of the codes used recently in an extensive test of cosmological
large-scale structure simulations~\cite{heitmann_05}. The cosmological
parameters were taken to be $\Omega_m=0.314$ (where $\Omega_m$
includes both dark matter and baryons), $\Omega_b=0.044$,
$\Omega_{\Lambda}=0.686$, $H_0=71$~km/s/Mpc, $\sigma_8=0.84$, and
$n=0.99$. These values are in concordance with the WMAP measurement
\cite{spergel_03}. The matter transfer function was generated using
the fits from Klypin \& Holtzman (1997), the initial redshift was
$z_{in}=50$, and periodic boundary conditions were used.


\section{Void Volume Function and Volume Fraction}
\label{sec:VolFunc}

The dark matter density field on a grid is obtained from the spatial
N-body particle distribution by applying the Cloud-In-Cell (CIC)
method followed by smoothing with a Gaussian filter. Using this
smoothed density field, we define a void as an individual region
bounded by one or more closed surfaces. Thus, a void is identified as
a set of grid sites with density contrast ($\delta \equiv (\rho -\bar{\rho})/\bar\rho$,
where $\rho$ and $\bar{\rho}$ are the density and mean density respectively)
 below a selected threshold ($\delta
\le \delta_c$) and satisfying a neighboring criterion consisting of
two conditions. First, that the six closest sites are the immediate
neighbors of the selected site and, second, that the immediate
neighbors of every neighbor are also the neighbors of the site.  The
volume of each void is estimated using the number of its sites
multiplied by the volume associated with the grid cell. All voids at a
given density threshold make up the void excursion set. We quantify
the density thresholds by the corresponding filling factors ($FF$),
\ie the fraction of the total volume in the excursion set. 

The void volume function $n(V)$ gives the number\footnote{It should be
noted that in the discussion below, the number of voids quoted is for
the particular simulation box volume $V_{\rm box}=256^3$ ($h^{-1}$Mpc)$^3$ and
the particular bin sizes: $\Delta V/V=1/\sqrt{2}$. The statistical
significance of the results can be directly assessed using the total
number of voids in every bin.} of voids having the given volume $V$
\be 
dN =n(V)dV. 
\ee 
The behavior of the void volume function is shown in the left panels
of Fig.~\ref{fig:gauss} for a density field at a very early epoch
$z=150$ -- essentially a Gaussian field -- whereas
Figs.~\ref{fig:number1} and \ref{fig:number2} show the void volume
function for evolved nonlinear dark matter density fields. These
results will be discussed in the following sections in detail. 

The volume fraction in voids, $\nu(V)$ specifies the amount of volume
in voids of a given volume $V$, following directly from the definition
of the void volume function 
$n(V)$ 
\begin{equation} 
\nu(V) dV =\frac{V}{V_t} n(V)\, dV, 
\end{equation} 
where $V_t$ is the volume of the simulation box. We stress that the
very largest void must be excluded from this statistic because of its
peculiar properties around the percolation transition, \eg Shandarin,
Sheth \& Sahni (2004). We will discuss this further below. The void
volume fraction is shown in the right panels of Figs.
\ref{fig:gauss}, \ref{fig:volume1} and \ref{fig:volume2} for the
Gaussian field and N-body simulations. 

\section{Gaussian Density Field} 
\label{sec:gauss}
We first briefly discuss voids in the case of a Gaussian density
field. As a cosmologically relevant example of such a field we
consider the initial particle density field used for the N-body 
simulations.  The code generating the initial conditions was used to
obtain ten realizations of a Gaussian field at the stage when
$\sigma_{\delta} \equiv <\delta^2>^{1/2}=0.0155$ (after smoothing at comoving $R_f=1\,\hm$),
which corresponds to an initial field at redshift $z=150$.
\begin{figure}
\epsfxsize= 8.4truecm\epsfbox{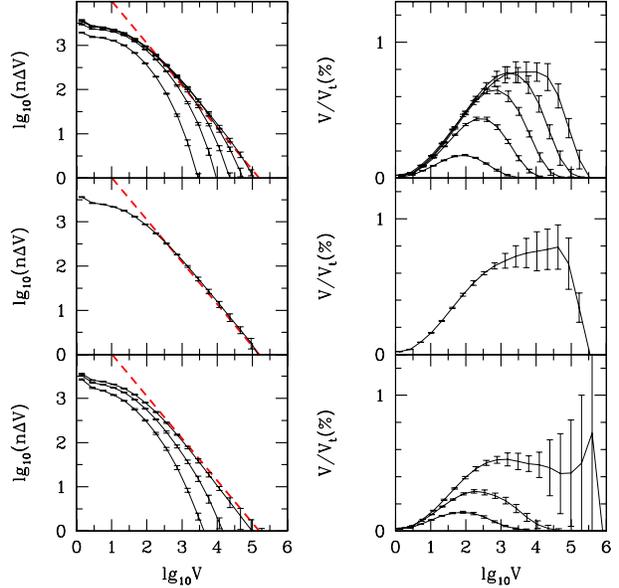} 
\caption{Gaussian field: $\sigma_{\delta}\approx 0.0155$. 
  The panels on the left show the void volume function and on the
  right the void volume fraction for different filling factors. The
  top panels show the both functions before the onset of percolation:
  $FF$=1\%, 3\%, 5\%, 7\% and 9\% (from bottom to top). The bottom
  panels show the curves after the onset of percolation: $FF$= 12\%,
  17\% and 25\% (from top to bottom ). The panels in the middle
  approximately correspond to the percolation transition; the curves
  correspond to $FF_c=10\%$ ($\delta_c=-0.02$). The mean and the error
  bars have been derived from ten realizations of the initial
  conditions. The dashed straight line in the left panels show the
  power law $n\Delta V = 8.9 \times10^{4} V^{-0.95}$, where volume $V$
  is measured in (h$^{-1}$Mpc)$^3$ and $\Delta V/V = 1/\sqrt{2}$.}
\label{fig:gauss}
\end{figure}

We first discuss the mean void function from the ten simulations. This
is displayed in the left panels of Fig.~\ref{fig:gauss}.  The top
panel shows the monotonic growth of the void function with increase of
the void filling factor from $FF=1\%$ (lowest curve) to $FF=$ 3\%,
5\%, 7\%, and 9\% (top curves) before the onset of percolation.  The
growth stops approximately at $FF=$10\% as shown in the middle panel.
This corresponds to the percolation transition: the merging of voids
results in the formation of a percolating network in the underdense
phase. After the onset of percolation the void volume function of
isolated voids monotonically decreases with increase in the filling
factor: the curves in the bottom panel correspond to $FF=$12\%, 17\%
and 25\% (from top to bottom).

A robust general scaling relation from percolation studies can now be
applied to our results. According to this relation~\cite{sta_aha_92},
the number of large voids scales as
\be 
n(V) \propto V^{-\tau}\exp(-cV),
\label{perc} 
\ee 
where $c \propto |FF-FF_c|^{1/\sigma}$; $FF<FF_c$ where $FF_c$ is the
filling factor at the percolation transition, and $\sigma$ and $\tau$
are real parameters. Our results for Gaussian fields are seen to be in
basic agreement with the percolation model with $\tau \approx 1.95$ as
shown by the dashed straight line in Fig.~\ref{fig:gauss}: at
percolation the void volume function is well-described by a power law
over three orders of magnitude. Voids with volumes in the range of the
scaling part of the void volume function would be referred to as ``critical
clusters''\footnote{In percolation theory, the term ``cluster'' refers
to a group of neighboring sites satisfying a criterion such as
``occupied'' or ``spin-up'', etc. In our case this criterion is set by
the density threshold $\delta_c$.} in percolation
theory because the power law behavior is characteristic of critical
phenomena and the properties of these voids are dominated by the
behavior at $FF_c$. The larger voids in the exponential tail of the
void volume function [Cf. Eqn.~(\ref{perc})] are rare and their 
properties are not controlled by the system behavior at $FF_c$.
The scaling properties of the excursion sets in Gaussian and especially 
nonlinear fields are interesting topics and deserve a more detailed 
study, however they are beyond the scope of the current work.
We will return to these issues in the future work.

The volume fraction is shown for the corresponding filling factors in
the right panels of Fig.~\ref{fig:gauss}.  The basic behavior of the
volume fraction is similar to that of the void function. The volume
fraction grows monotonically as the filling factor is increased until
the onset of percolation (right top panel), after which it
monotonically decreases (right bottom panel).  The onset of
percolation is analogous to a phase transition and therefore, due to
the expected increase in fluctuations near a ``phase change'', it is
not surprising that the error bars are significantly greater in the
vicinity of $FF_c$.  Although the statistical mean for the percolation
transition is $FF_c =10\%$, this number will change from realization
to realization. When analyzing a particular realization, the
percolation threshold for the particular realization should be used
rather than the statistical mean value. As a consequence of this,
somewhat different values for the percolation threshold than given in
this section will be encountered in the following analysis. The
difference is small, $\sim 0.5-1\%$, well inside the $1\sigma$ range.

\section{Voids in N-body Simulations}
\label{sec:n-body}
\begin{figure}
\epsfxsize= 8.4truecm\epsfbox{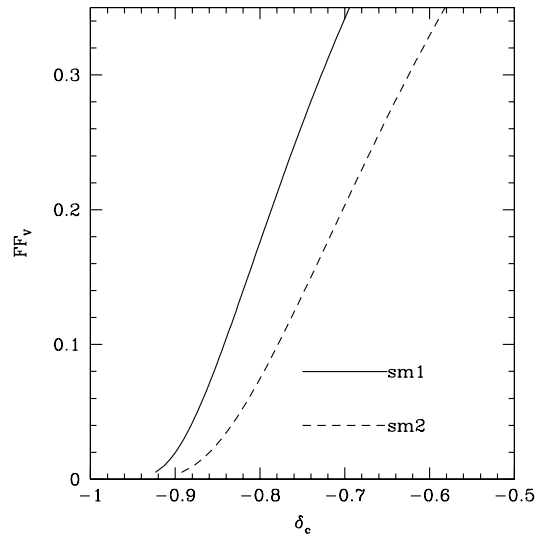} 
\caption{Void filling factor $FF_v$ versus density contrast 
  threshold $\delta_c$  at $z=0$ for two smoothing scales: sm1 and sm2
  corresponding to $R_f=1\hm$ and $R_f=2\hm$ respectively. The
  smallest filling factor is 1\% and the largest 35\%. }
\label{fig:FF_del}
\end{figure}

In this section we present a comparative analysis of voids in the
initial ($z=50$) and final ($z=0$) density fields taken from the
N-body simulation. This analysis is targeted at establishing which
properties of voids are inherited and which are acquired in the course
of evolution. Figure \ref{fig:FF_del} relates the filling
factor to the density contrast. As we shall see later the percolation thresholds
at $z=0$ incidentally are close to $\delta=-0.8$ (-0.78 and -0.75 
for $R_f =1\hm$ and $R_f =2\hm$ respectively)  corresponding to the density 
contrast of the spherically symmetric top-hat void model when  it reaches the
first shell crossing. At these thresholds about a half of the volume of the 
underdense excursion set is already in one percolating void.

\begin{figure}
\epsfxsize= 8.4truecm\epsfbox{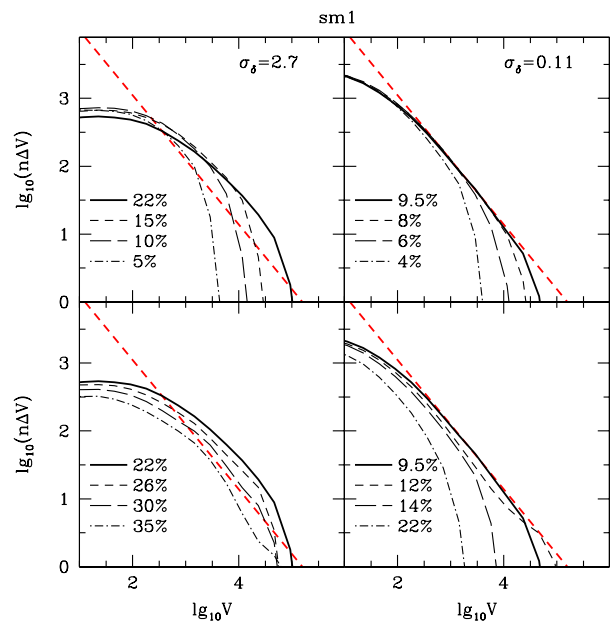} 
\caption{Void volume function $n(V)\Delta V$ with the density field
  smoothed at $R_f=1\hm$. The panels on the left are for $z=0$ and on
  the right, $z=50$. The top panels show the void volume functions at
  small filling factors before percolation occurs ($FF_c=$22\% ,
  $\delta_c=-0.78$ at $z=0$ and $FF_c = 9.5\%$, $\delta_c=-0.14$ at
  $z=50$) and the bottom panels after percolation.  The heavy solid
  lines in the top and bottom panels are the same; they show the void
  volume function approximately at percolation. The dashed straight
  line is the same as in Fig. \ref{fig:gauss}, i.e., the power law
  $n\Delta V=8.9\times 10^4V^{-0.95}$ characteristic of the initial
  Gausssian density field.}
\label{fig:number1}
\end{figure}
\begin{figure}
\epsfxsize= 8.4truecm\epsfbox{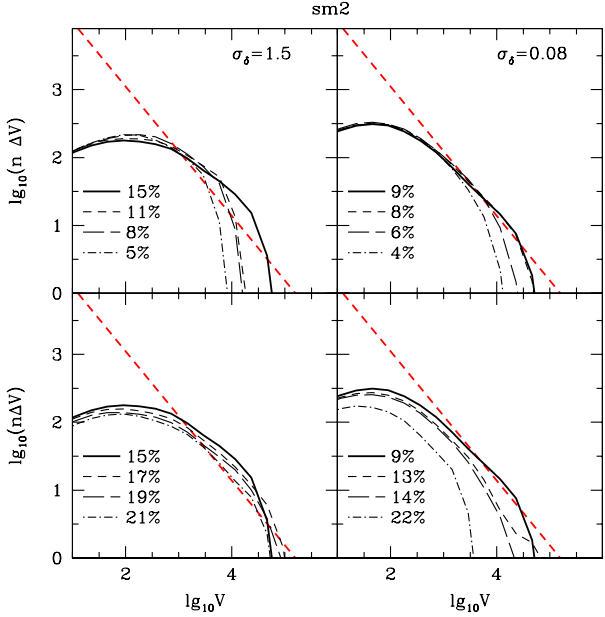}
\caption{Void volume function $n(V)\Delta V$ with the density field
  smoothed at $R_f=2\hm$.  The critical parameters are: $FF_c=$15\% ,
  $\delta_c=-0.75$ at $z=0$ and $FF_c = 9\%$, $\delta_c=-0.10$ at
  $z=50$.  The organization of the figure is the same as in
Fig.~\ref{fig:number1}.} 
\label{fig:number2}
\end{figure}

It has been demonstrated in previous studies~\cite{sh_she_sah_04} that
the void volume function is dominated by small voids. This is not
surprising at the linear stage of evolution of the density field, when
the field is essentially Gaussian, but it is also true in the
nonlinear regime. This is illustrated by Figs.~\ref{fig:number1} and
\ref{fig:number2} that show the void volume function at different
filling factors and smoothing scales. In these figures, results are
shown for the same underlying density field but with different
Gaussian filtering windows, $R_f=1\hm$ and $R_f=2\hm$
respectively\footnote{We also have analyzed the fields smoothed with
$R_f=4\hm$ but do not show the results here because they do not reveal
anything significant.}. 

In Figs.~\ref{fig:number1} - \ref{fig:number2}, the two panels on the
left show the number of voids at $z=0$ while the two panels on the
right represent the initial density field at $z=50$. The top panels
show the void volume functions at several filling factors
corresponding to relatively low density thresholds when voids do not
percolate. The bottom panels are complementary to the top panels and
show the void volume function at the thresholds {\em after}
percolation.  The heavy solid lines are the same in the top and bottom
panels and show the void volume function approximately at the
percolation thresholds. These thresholds correspond to a filling
factor, FF=9-9.5\%, for the linear fields regardless of the
smoothing scale, and decreases from 22\% at the smallest smoothing scale
($R_f=1\hm$) to 14\% at $R_f=2\hm$ for the evolved field. The dashed
straight line is the power law obtained previously for a Gaussian
field with the same cosmological parameters.

Figure \ref{fig:number1} clearly demonstrates that the evolution
results in a decrease of the number of small voids and an increase of
the number of large voids, with the approximate transition boundary
between the two behaviors at $V\sim 300 \hm$.  The field smoothed
with $R_f=2\hm$ (Fig. \ref{fig:number2}) yields qualitatively similar
results, however, there are some quantitative differences. Relative to
the $R_f=1\hm$ smoothing, the number of small voids is lower in both
the initial and final fields, as well as the number of large
voids. The former is not unexpected because smoothing with greater scale
erase smallest voids.  However, the latter result seems to be less
obvious since the smoothing scale is relatively small and seemed not to 
influence large voids. Probably the explanation is in the significant reduction of the
percolation filling factor in more smoothed density field that makes voids
less isolated. As a result they merge into the percolating voids before
they reach large volumes.

The void volume function at the fully developed nonlinear stage $z=0$
is clearly not a power law, while the initial field ($z=50$) is mostly
in good agreement with the scaling discussed in the previous
section. The lack of the scaling (power law) component in the void
volume function at the nonlinear stage is a possible indicator of the
presence of large-scale coherence in the density field. This issue
will be addressed elsewhere.
 
In agreement with the Gaussian case, the results displayed in
Figs.~\ref{fig:number1} and \ref{fig:number2} show that the number of
voids monotonically increases with the filling factor and reaches a
maximum at the percolation threshold (top panels), then monotonically
decreases with the filling factor (bottom panels).  We conclude that
the largest number of voids -- practically of every volume -- is
reached approximately at the percolation thresholds.  Below we show
that a small number of the largest voids make up most of the volume of
the void excursion set while the majority of voids have very small
volumes and constitute only a small fraction of the void excursion
set.  This is true for all filling factors at all smoothing
scales~\cite{sh_she_sah_04}.

\begin{figure}
\epsfxsize= 8.4truecm\epsfbox{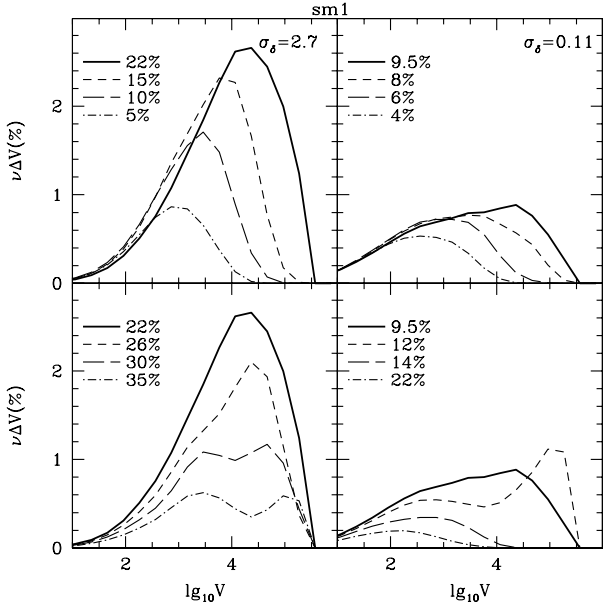} 
\caption{Void volume fraction $\nu(V) \Delta V$ with density field
smoothing of $R_f=1\hm$, following Fig.~\ref{fig:number1}.}
\label{fig:volume1}
\end{figure}

\begin{figure}
\epsfxsize= 8.4truecm\epsfbox{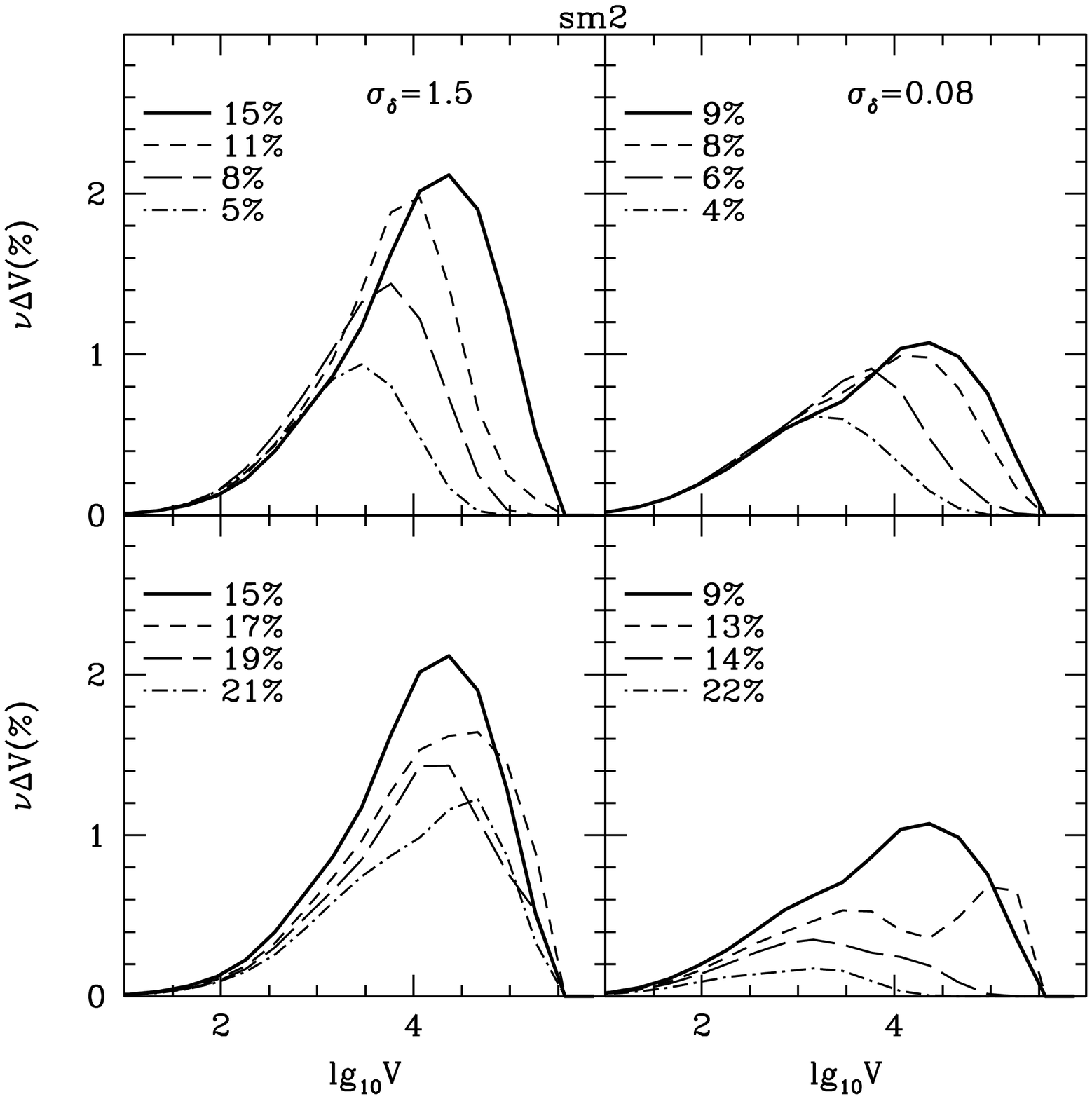} 
\caption{Void volume fraction $\nu(V)\Delta V$ with density smoothing
scale of $R_f=2\hm$, following Fig.~\ref{fig:number2}.}
\label{fig:volume2}
\end{figure}

Figures ~\ref{fig:volume1} and \ref{fig:volume2} demonstrate -- in a
manner similar to the void volume function
(Figs.~\ref{fig:number1}-\ref{fig:number2}) -- that the void volume
fraction grows monotonically with the filling factor until it reaches
the percolation threshold shown by the heavy solid lines (top panels).
Thereafter, it monotonically decreases with the filling factor (bottom
panels). This occurs because the largest void becomes the percolating
void and contains most of the volume of the void excursion set.  The
volume fraction of voids reaches a maximum at the percolation
thresholds for both linear and nonlinear fields.


Comparing the void volume fraction (Figs.~\ref{fig:volume1} and
\ref{fig:volume2}) with the void volume function
(Figs.~\ref{fig:number1} and \ref{fig:number2}) we find that a
relatively small number (less than roughly $\simeq 10\%$) of the
largest voids [with approximately $V \ge 10^3 (\hm)^3$] make up most
(roughly $\sim 90\%$ or even more) of the void excursion set (see also
Shandarin, Sheth \& Sahni 2004). In their numbers small voids
dominate, but they do not contribute significantly to the entire
volume of the void excursion set.

\begin{figure}
\epsfxsize= 8.4truecm\epsfbox{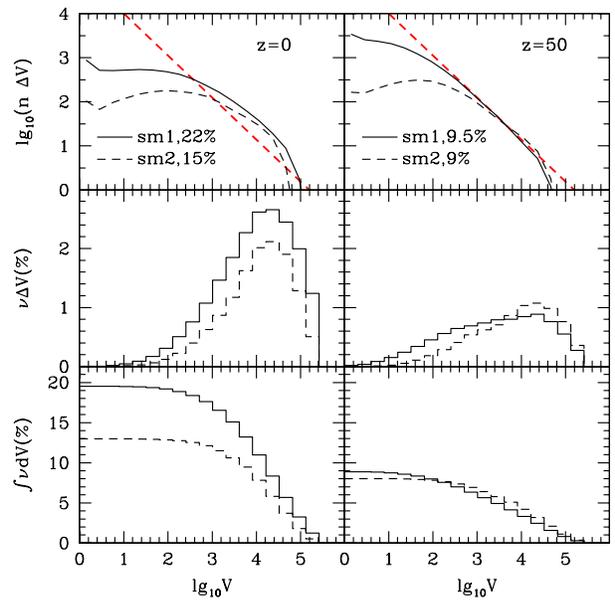} 
\caption{Summary of the information in
Figs.~\ref{fig:number1}-\ref{fig:volume2}; the volume functions and
volume fractions are shown at the corresponding percolation thresholds.
Top: the volume functions at $z=0$ (left) $z=50$ (right). 
Middle: the corresponding volume fractions.
Bottom: the cumulative void fractions:  $\int_V^{\infty} \nu dV$}
\label{fig:combined}
\end{figure}

The results of this section are summarized in Fig.~\ref{fig:combined}.
The top panels show the void volume functions, middle panels, the
volume fraction and bottom panels, the cumulative volume fraction at
the percolation thresholds respectively. Solid lines correspond to the
filtering scale $R_f=1\hm$ and dashed lines to $R_f=2\hm$.  We
conclude that the percolation threshold is a natural choice
characterized by the largest numbers of voids as well as the largest
volume fraction in the void excursion set (compare to Shandarin, Sheth
\& Sahni 2004). As mentioned already, a not completely obvious result
is that smoothing the density field reduces the number of large voids
along with the expected reduction of the number of small voids. It
appears that this effect is smaller or even absent in Gaussian fields,
however, this may be a statistical fluctuation due to insufficient statistics.

\begin{figure}
\epsfxsize= 8.4truecm\epsfbox{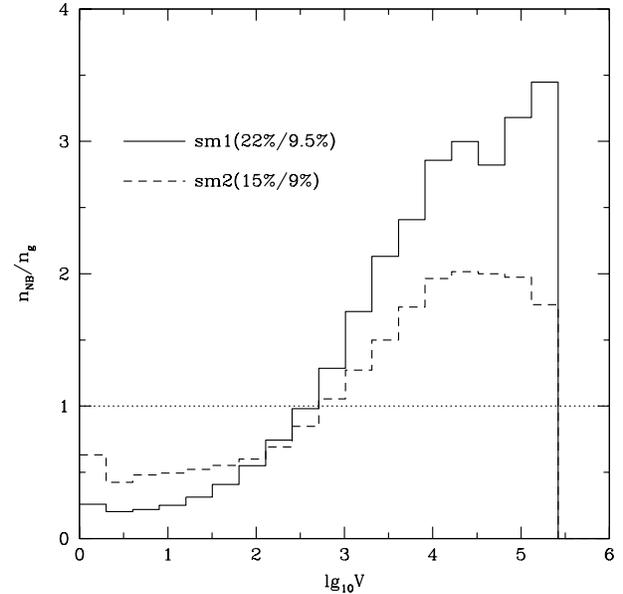} 
\caption{Ratio of number of voids in the N-body simulations to the
  number of voids in the initial density field at $z=50$ as a function
  of volume.  The voids were identified at the corresponding
  percolation thresholds in every density field: for N-body simulation
  at $FF=$22\% and 15\% at $R_f=1\hm$ and $R_f=2\hm$ respectively, and
  for initial state at $FF=$9.5\% and 9\% at these smoothing scales
  respectively. }
\label{fig:ratio}
\end{figure}

The major effect of the nonlinear evolution is summarized in
Fig.~\ref{fig:ratio} where the ratio of the number of voids in the
N-body simulations to the number of voids in the initial density field
at $z=50$ is plotted as a function of void volume. Voids smaller than
$10^3 (\hm)^3$ are strongly suppressed while voids with greater
volumes become approximately three times more abundant than in the
parent density field. As a result, the voids smaller than roughly
$10^3(\hm)^3$ comprise less than $\sim 10\%$ of the void excursion set
(the bottom panels of Fig.~\ref{fig:combined}). Consequently, we
concentrate mostly on larger voids in our further analysis. A somewhat
counterintuitive result is that very large voids [$V > 3\times10^5
(\hm)^3$] are apparently not more abundant at the nonlinear stage than
in the initial density field; this could be an artifact due to the
finite size of the simulation box, however, and should be checked
using larger volume simulations.

We now discuss how the excursion set definition of voids -- as regions
bound by isodensity surfaces -- results in seeming disagreement with
other definitions of voids. Most significantly, we find that the
amount of space in isolated voids is not large. For a smoothing scale
of $R_f=1\hm$ it is only about 20\% of all space, and for $R_f=2\hm$
it is even less: about 15\%. In contrast, other studies would claim
that most of space is taken up by voids. This discrepancy is rather
easy to explain: We deal with isolated voids naturally defined by
isodensity surfaces, while in other studies voids are built using
procedures that follow the density contours only approximately,
thereby allowing closure of the boundaries of the constructed voids by
non-isodensity surfaces. With our definition of voids one can get as
much of void space as one wants by raising the density threshold, but
then the natural isodensity surfaces would inevitably build a single
void percolating throughout the entire volume plus a few small
isolated voids. Cutting the percolating void into pieces would finish
the job of constructing large ``isolated'' voids, however this brings
a certain arbitrariness, and therefore bias, into the analysis.  Other
algorithms actually suggest various methods of doing this, but without
explicit discussion. Our procedure does not cut the largest void --
neither before nor after it percolates -- and therefore the voids
found by it can make up only a relatively small fraction of space. At
even greater density thresholds the percolating void is absolutely
dominant~\cite{sh_she_sah_04}.  Reducing the smoothing scale increases
the number of voids and the amount of space occupied by isolated voids
(Fig. \ref{fig:combined}).  However, if a uniform filter is used for
smoothing then the intrinsic discreetness of N-body simulations takes
over and prevents further reduction of the smoothing scale. An
adaptive smoothing filter would help to ameliorate this problem.

\section{Fitting to ellipsoids}
\label{sec:fitellipsoid}

\begin{figure*}
\epsfxsize= 16truecm\epsfbox{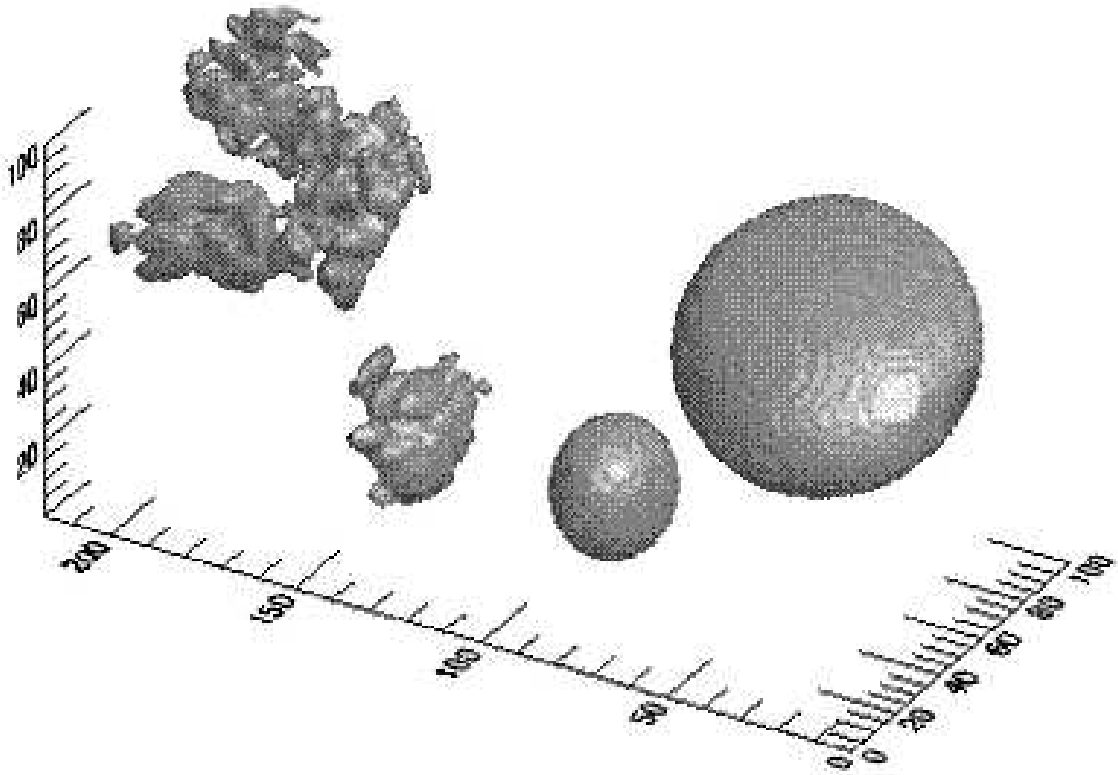} 
\caption{Comparison of an ellipsoid fit to high and low porosity voids. 
The low porosity small void (bottom left) has volume of $V=7500$ cells, 
Porosity $P=V_{E}/V_{V}=1.92$, and Inverse Porosity $IP=V_{V}/V_{E}=0.52$; 
the fitting ellipsoid is at the bottom right. The high porosity large void 
(top left) has volume of $V=24000$ cells, $P=4.72$, and $IP=0.21$.}
\label{fig:h_l_por_voids}
\end{figure*}

As discussed in the Introduction, the actual shapes of voids are often
taken to be approximately spherical or ellipsoidal. In order to
investigate possible systematic difficulties with this kind of
fitting, we identify all voids at a selected density threshold
(roughly at the percolation threshold) and then fit each void by an
ellipsoid having the same inertia tensor as the void itself:
\begin{eqnarray}
&&J_{xx}=\sum_i m_i(y_i^2+z_i^2),\qquad  J_{xy}=-\sum_i m_i x_i y_i,\\
&&\qquad
\mbox{plus cyclic permutations},\nonumber
\end{eqnarray}
where $m_i$ and $x_i, y_i, z_i$ are the mass and coordinates of the
site with respect to the void center of mass and the sum is taken over
all sites belonging to the void. The semiaxes of the fitting ellipse
can be found from the principal moments of inertia $J_1, J_2, J_3$,
viz., 
\begin{equation} 
a^2=\frac{5}{2M}(J_2+J_3-J_1) \qquad \mbox{plus cyclic
  permutations}, 
\end{equation}
where $M=\sum m_i$ is the mass of the void. Our focus here is on the
geometrical properties of voids and therefore we compute the inertia
tensor of an empty void by setting all the $m_i$ to a constant.  We
reserve the study of another interesting option, where the $m_i$ are
actual dark matter masses in the grid cells, for future
work. Figure~\ref{fig:h_l_por_voids} illustrates the fitting procedure
outlined above for two example voids: in the figure, the voids are
shown on the left with the corresponding ellipsoids on the right.
The axes of the fitting ellipsoid, $a>b>c$, can be used to evaluate
the volume of the ellipsoid $V_E=(4\pi/3)abc$ as well as the ratios
$b/a$ and $c/b$ which approximately characterize the shape of the void. 
The Planarity $P=(1-b/a)/(1+b/a)$ and Filamentarity
$F=(1-c/b)/(1+c/b)$~\cite{sah_sat_sh_98,sh_she_sah_04} are equivalent
measures of void shape. 

If a void is ideally ellipsoidal then its volume must equal that of
the fitting ellipsoid. Actual voids are not ellipsoidal in shape,
however~(Fig.~\ref{fig:h_l_por_voids}), thus their volumes $V_V$ are
always smaller then the volumes of the fitting ellipsoids $V_E$.  The
ratio $P=V_E/V_V$ -- dubbed the Porosity -- quantifies the quality of
the ellipsoidal fit.  We use the Inverse Porosity $IP = V_V/V_E$ ($0
\le IP \le1$) as a simple estimate of the goodness of the ellipsoidal
fit.  If $IP$ is close to unity, the fit may be regarded as good,
otherwise the smaller the $IP$, the worse the fit.

\section{Sphericity}
\label{sec:sphericity}

\begin{figure}
\epsfxsize= 8.4truecm\epsfbox{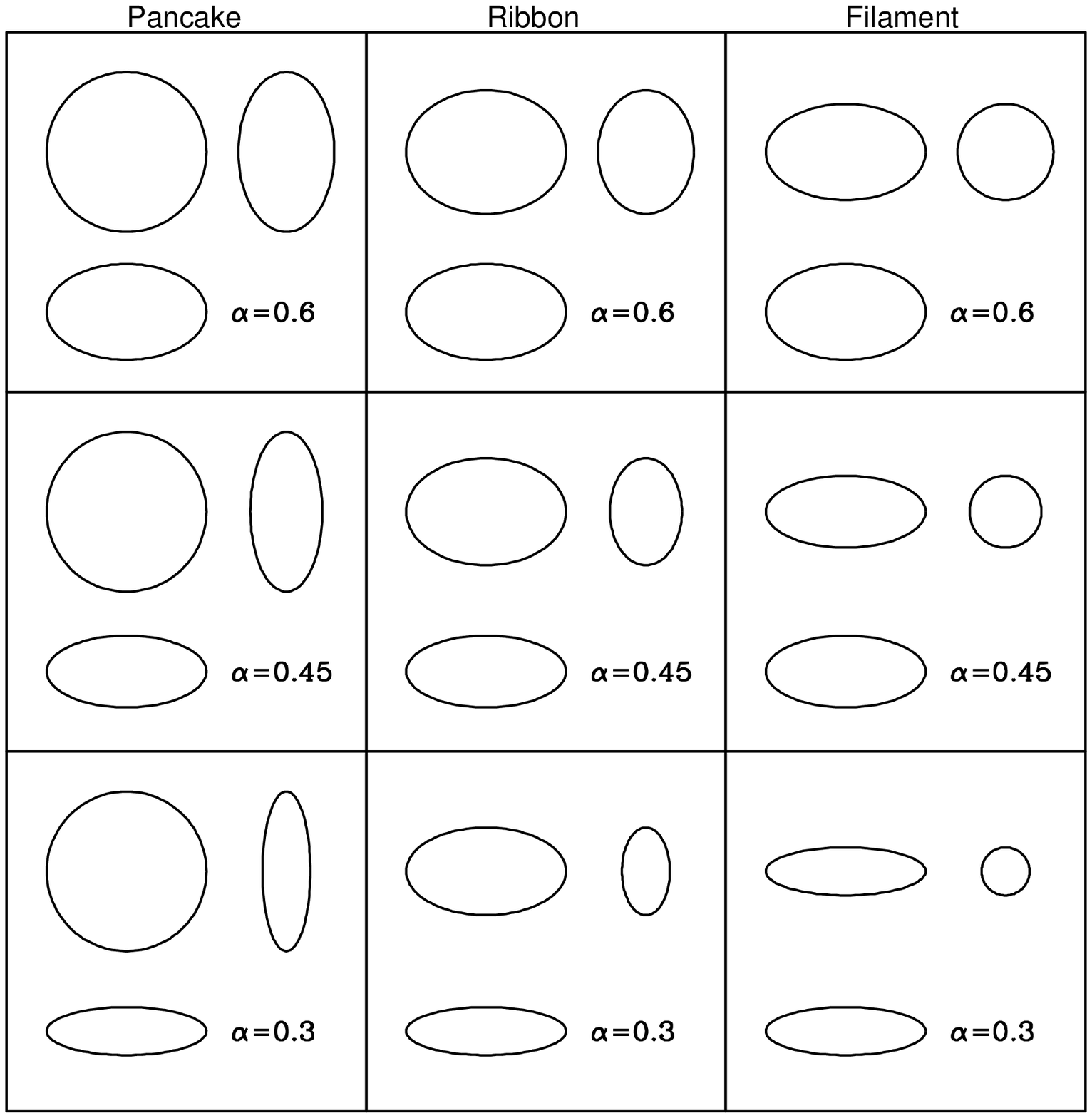} 
\caption{Three principal projections of an ellipse are shown in every
panel. Each column shows one of three archetypes of the deformed
sphere parameterized by one number $\alpha=c/a$ ($0\le \alpha \le 1$):
pancake-like ($b/a=1, c/b =\alpha$), ribbon-like ($b/a=\sqrt{\alpha}, 
c/b=\sqrt{\alpha})$, and filament-like ($b/a=\alpha, c/b=1$).
}
\label{fig:sphericity}
\end{figure}

Deforming a sphere into an ellipsoid with three axes, $a\ge b\ge c$,
one obtains a two-parameter family of shapes that can be fully
characterized by two ratios: $b/a$ and $c/b$. Nevertheless, we would
like to approximately describe the departure from a sphere with only
one parameter. Ideally the parameter should have a simple geometric
interpretation and only a small variation when the remaining degree of
freedom is varied. We suggest the use of the ratio of the smallest to
largest axis, $\alpha=c/a$, for this purpose. Using this, one can
distinguish three archetypal shapes among the entire two-parameter
family characterized by $\alpha$: pancake-like with $b/a=1$ and $c/b=\alpha$,
filament-like with $b/a=\alpha$ and $c/b=1$, and ribbon-like with
$b/a=c/b=\sqrt{\alpha}$. All three have very distinct limits at
$\alpha \rightarrow 0$. Figure~\ref{fig:sphericity} provides a visual
illustration of the variation of shapes from a pancake through a
ribbon to a filament for three values of $\alpha$.

\section{Shapes of voids in N-body simulations}
\label{sec:shape}

In this section we discuss the shapes of voids in the simulated
$\Lambda$CDM model at the present epoch ($z=0$). As earlier, we
analyze the density distribution in real space at three different
smoothing scales: $1h^{-1}, 2h^{-1}$, and $4h^{-1}$Mpc.

\subsection{Mean values}
\label{subsec:mean}

\begin{figure}
\epsfxsize= 8.4truecm\epsfbox{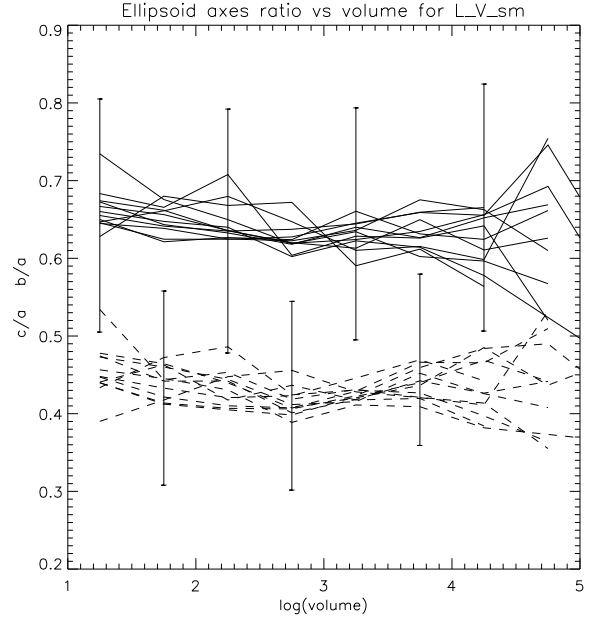}
\caption{
Axis ratios as a function of volume. The average of the axis ratios is
independent of the filling factor and smoothing length. Except for
very small voids, the average is constant with volume. 
}
\label{fig:axes-volume}
\end{figure}

The mean Sphericity as a function of the volume is shown in
Fig.~\ref{fig:axes-volume}. Every line shows the ratio at a different
filling factor and for all three smoothing scales. The variance is
shown by vertical bars for one of the lines, the others being very
similar. It is clear that there is no significant statistical
difference between the mean shapes of voids at different filling
factors or smoothing scale: on average $b/a\approx 0.65$ and
$c/a\approx0.45$. Thus, if a region of the underdense excursion set
is approximated by an ellipsoid then the ellipsoid should be triaxial with
the shortest axes at least twice shorter than the longest one. A substantial
number of ellipsoids has a shortest axes as small as a third of the longest
one.
Figure \ref{fig:sphericity} provides the visual illustration of such
ellipsoids.

\begin{figure}
\epsfxsize= 8.4truecm\epsfbox{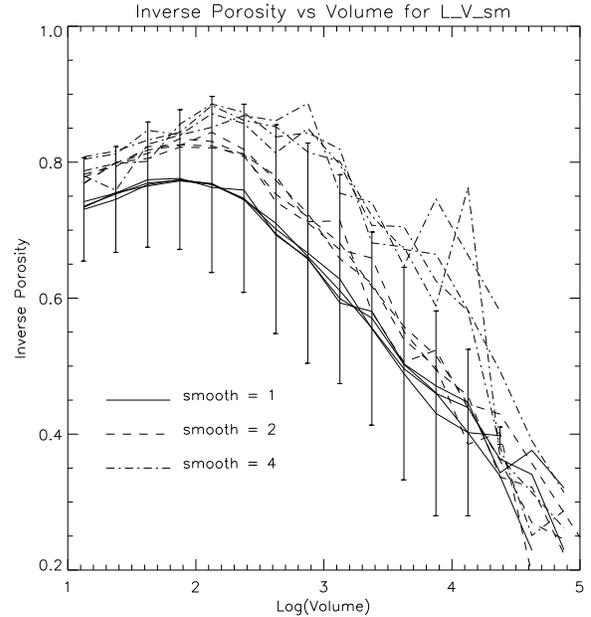}
\caption{
  Inverse Porosity ($IP=V/V_\ell$) as a function of void volume.
  The larger the volume the larger is the Porosity on average. There
  is little dependence on the filling factor, while the smoothing
  length systematically suppresses the Porosity.  }
\label{fig:porosity-vol}
\end{figure}

In Fig.~\ref{fig:porosity-vol}, we illustrate the dependence of the
mean Inverse Porosity $IP = V_V/V_E$ as a function of the volume of
the void for different filling factors and smoothing scales. Following
the behavior of the Sphericity, the Porosity does not show a strong
dependence on the filling factor, however there are two other
characteristics that are very different from the mean Sphericity. The
first is a systematic growth of the mean Inverse Porosity with the
smoothing length, albeit with large variance. This is an obvious
consequence of smoothing; the voids in the smoothed density fields
being more ellipsoidal. The other is a clear dependence on the volume
of the voids: the larger the volume, the smaller the Inverse Porosity
and therefore the worse the ellipsoidal fit. Combining the information
in Figs.~\ref{fig:porosity-vol} and \ref{fig:combined} we conclude
that large voids $V > 3\times10^3\hm^3$ -- making most of the void
excursion set -- cannot generically be adequately approximated by
ellipsoids.

\subsection{Distribution Functions}
\label{subsec:distfun}

\begin{figure}
\epsfxsize= 8.4truecm\epsfbox{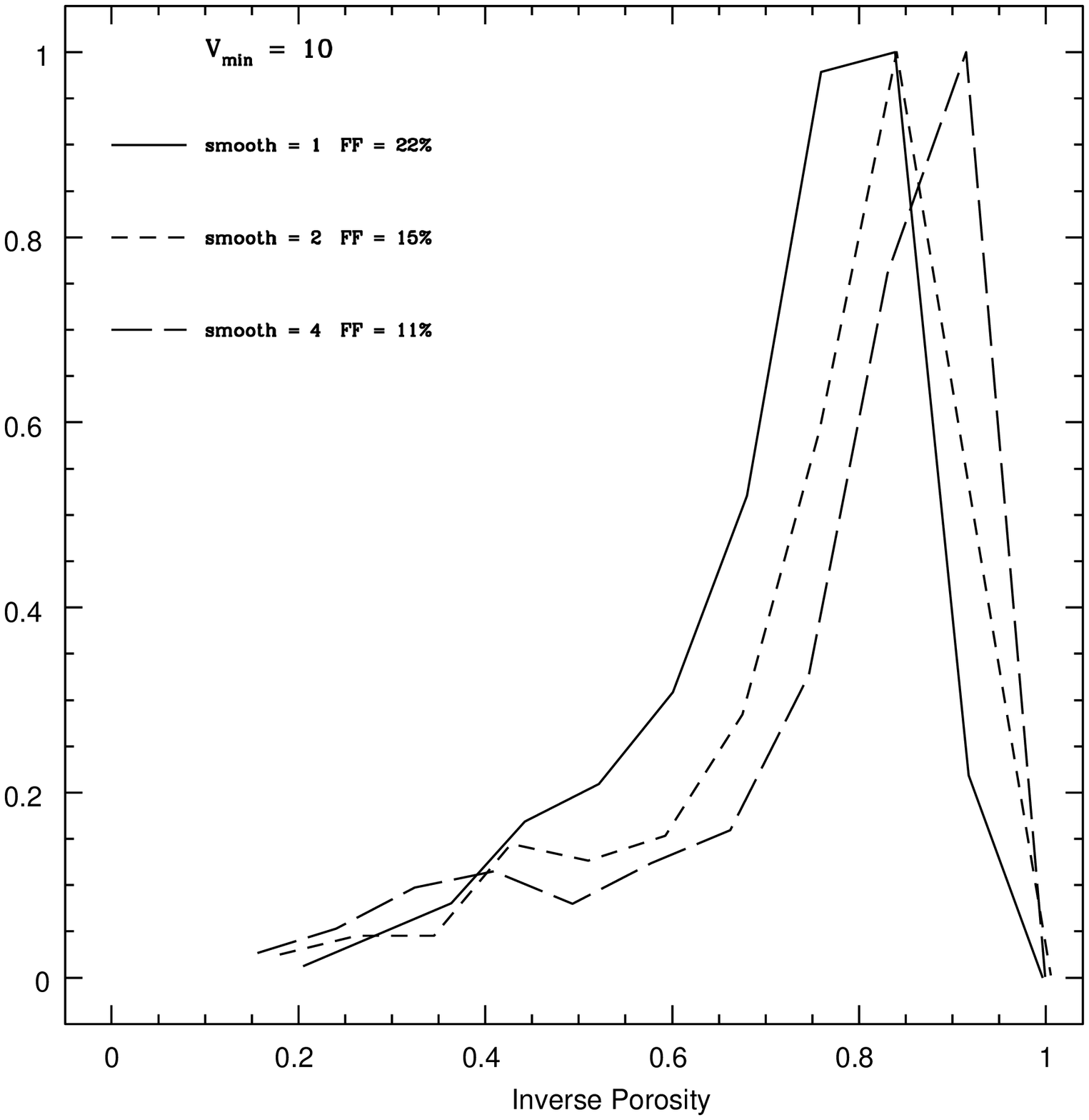} 
\caption{Normalized distribution functions of the Inverse Porosity 
  ($V_V/V_E$) of voids with volumes $V_V > 10 (\hm)^3$ at three
  smoothing scales.  The distribution functions are shown for the
  corresponding percolation thresholds.  }
\label{fig:poros_10}
\end{figure}

\begin{figure}
\epsfxsize= 8.4truecm\epsfbox{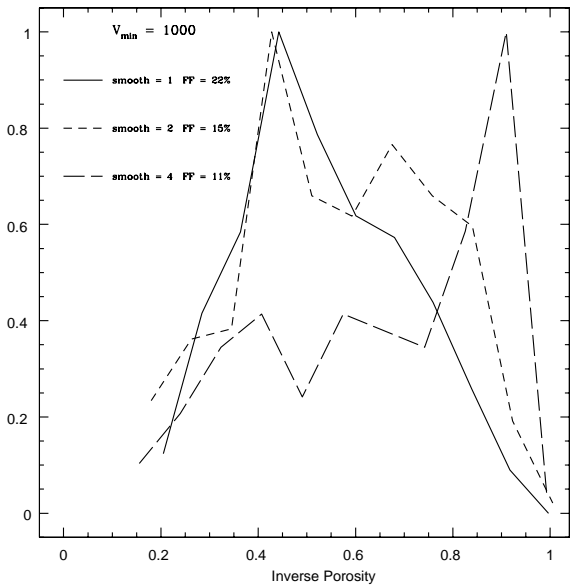} 
\caption{Normalized distribution function of the 
  Inverse Porosity ($V_V/V_E$) of voids with volumes $V_V > 1000
  (\hm)^3$ at three smoothing scales following Fig.~\ref{fig:poros_10}.}
\label{fig:poros_1000}
\end{figure}

Statistical information regarding the quality of elliptical and
spherical void fits can be gleaned from the distribution functions for
the Inverse Porosity and the Sphericity. We begin by first discussing
the Inverse Porosity distribution function.
Figures~\ref{fig:poros_10} and \ref{fig:poros_1000} show the IP
distribution for density fields smoothed on three scales: $R_f=1\hm$
(solid line), $R_f=2\hm$ (short-dashed line), and $R_f=4\hm$
(long-dashed line). Figure~\ref{fig:poros_10} includes all the voids
with volume greater than $10(\hm)^3$ and Fig.~\ref{fig:poros_1000}
those with volume greater than $1000(\hm)^3$. The first cut is imposed
because the shape measurement of small voids $V<10 (\hm)^3$ is not
reliable. The second cut selects only large voids $V>1000(\hm)^3$
comprising approximately 90\% of the void excursion set
(Cf.~Fig.~\ref{fig:combined}). If voids with $V<1000 (\hm)^3$ are
included, the IP distribution peaks relatively close to unity,
implying that elliptical fits are quite good for the majority of the
voids. As we demonstrated earlier, however, small voids dominate only
in number and not in volume. If small voids are excluded
(Fig.~\ref{fig:poros_1000}) then the IP distribution peaks at 
$V_V/V_E \approx 0.45$ for $R_f=1\hm$ and $R_f=2\hm$. In this case the
ellipsoidal fit is quite poor for most of the larger voids that make
up roughly 90\% of the void excursion set. Therefore, in the following
analysis we will exclude voids with volumes less than $1000 (\hm)^3$.


We note that the IP distribution in Fig.~\ref{fig:poros_1000} displays
a trend of the distribution peak shift to the right with increased
smoothing scale, with an especially pronounced change for
$R_f=4\hm$. This is expected as smoothing will improve the ellipsoidal
fit for the smallest voids considered; for $R_f=4\hm$, a sphere with
the radius of the smoothing length has a volume of $270 (\hm)^3$ which
is more than one quarter of the volume of the smallest voids that
dominate the void number distribution.

\begin{figure}
\epsfxsize= 8.4truecm\epsfbox{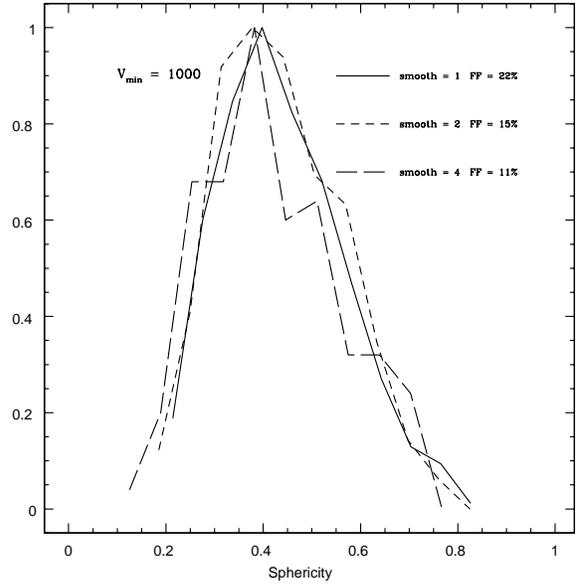} 
\caption{Sphericity distribution function of voids with volumes
$V_V > 1000 (\hm)^3$  at three smoothing scales. The distribution
functions are shown for the corresponding percolation thresholds.
}
\label{fig:spher_1000}
\end{figure}

Turning to the Sphericity, Fig.~\ref{fig:spher_1000} shows the
Sphericity distribution function for large voids $V_V > 1000
(\hm)^3$. In this case, the results from fields smoothed on different
scales are not very different.  The mean Sphericity of the fitting
ellipsoids is about 0.45 (in agreement with
Fig.~\ref{fig:axes-volume}).  The middle panels of
Fig.~\ref{fig:sphericity} show three archetypal ellipsoids
corresponding to this value. The top and bottom panels show the range
of Sphericity for most of the voids.

\section{Summary and Discussion}
\label{sec:discussion}

In this paper, we studied the distribution of underdense regions
(voids) in N-body simulations for the $\Lambda$CDM model, voids being
defined as individual regions of the underdense excursion set: $\delta
< \delta_c$. In practice, we additionally applied the neighbor of
neighbor criterion for grid sites satisfying the above threshold
condition. By defining the void volume function, we then showed that
the largest number of voids of any size is reached just before the
onset of percolation. In other words, as the density threshold is
increased more and more voids are found and the total volume fraction
in voids increases; this continues up to a critical point beyond which
the number of voids and the volume fraction in all but the percolating
void begin to decrease.

The percolation transition motivates a particular density threshold
and corresponding filling factor to focus further analysis, i.e., the
percolation threshold itself. After percolation, though the volume of
the underdense excursion set continues monotonically to increase with
the growth of the density threshold, the number and volume of
individual voids decrease because merging of individual voids into the
single percolating void dominates. The percolating void continues to
grow monotonically with the threshold, mostly by absorbing individual
voids. Choosing a particular threshold above the percolation point
seems to be arbitrary in most cases and may significantly affect the
results one obtains.

One could cut the percolating void found at an arbitrary threshold
into pieces. But without any clear physical motivation this would
introduce an additional arbitrariness which we wished to avoid. It
should also be noted that special precaution should be exercised in
one's notion of a void, as it is always possible to design an
algorithm producing more or less spherical voids if so desired. The
guideline provided by Icke (1984) suggesting that voids become more
spherical in the course of dynamical evolution should be reconsidered
in a more realistic environment rather than in the case of a single
void placed in a homogeneous universe. The external shear is likely to
play an important role~\cite{eis-loe-95} and must be included in the
model. Even in the case of a single void the effect may be
quantitatively insufficient~\cite{ber_85}. We will address these
issues in future work.


As a cautionary note we stress that the reported results may
depend on the smoothing procedure adopted here and the values of the
filtering scale. This is an issue with essentially all proposed
definitions of voids and needs to be further investigated.

The density field at an early enough epoch is essentially Gaussian
random on scales below the smoothing scales employed here. We
considered ten realizations of a density field at $z=150$ and showed
that the behavior of the void volume function closely followed a
mathematical form motivated by percolation theory; at later redshifts,
the field is no longer Gaussian and this form does not hold.

At the present epoch, we found that less than 10\% or so of the
largest voids, with volumes greater than about $10^3 (\hm)^3$, hold
more than 90\% of the void excursion set while small voids dominate in
their numbers. The dynamic evolution of voids results in a significant
decrease of the number of small voids [$V<10^3 (\hm)^3$] and an
increase in the number of large voids.

In order to study effective void shapes, we fit the underdense regions
to ellipsoids with the same inertia tensor as the voids. We used this
to define the Sphericity and (Inverse) Porosity of the voids and
showed that most large voids are not spherical and have irregular
geometries which lead to high porosity (low Inverse Porosity). The
prevailing assumption of spherical voids, used extensively in the
literature, is not a good approximation if the actual isodensity
surfaces are employed as the void boundaries. We found that the ratios
of the semi-major axes have mean values of $0.65$ and $0.45$ and the
Sphericity has a mean value of $c/a \approx 0.45$, (where $c$ and $a$
are the smallest and the largest semiaxes of the fitting ellipsoids),
which is clearly not spherical. In addition, the Porosity measure
showed that large voids are quite porous whereas small voids are more
regular, with even ellipsoidal fitting becoming problematic for the
larger voids.

Two questions are of potential relevance here. First, the representing
ellipsoid has the same inertia tensor as a homogeneous void while
actual voids are inhomogeneous. Therefore one could ask: would the
true density profile change the fitting ellipsoid to be closer to a
sphere? The obvious answer is no; it is natural to expect the central
part of a void to be more spherical than the boundary region but this
would have a smaller effect on the inertia tensor since the density at
the center of the void is lower than near the boundary.

The second question relates to the shapes of voids before the onset of
percolation. Is it possible that the percolating void has the shape of
a set of spheres connected by narrow tunnels -- one of the major
concerns of Kauffmann \& Fairall (1991) and El-Ad \& Piran (1997)?
This appears unlikely because the percolating void is made by the
merger of individual voids as the density threshold is increased. The
merger process happens quite quickly near the threshold (\ie with a
small change of the threshold) therefore the individual voids making
up the percolating void cannot change their shapes appreciably. Our
results show that the largest of them are neither spherical nor even
elliptical. Additionally, study of individual voids before and after
percolation do not show much difference~\cite{sh_she_sah_04}.

\noindent{\bf Acknowledgments}
 
S.H. and K.H. acknowledge support from the Department of Energy via
the LDRD program of Los Alamos National Laboratory. The calculations
described herein were performed primarily using the computational
resources of Los Alamos National Laboratory. A special acknowledgment
is due to supercomputing time awarded to us under the LANL
Institutional Computing Initiative. This research is supported by the
Department of Energy, under contract W-7405-ENG-36. H.A.F. was
supported in part by the GRF at the University of Kansas. S.S. was
partly supported by NSF-RP087 grant.  S.S., K.H., and S.H. acknowledge
support from the Aspen Center for Physics where the manuscript was
finalized. We acknowledge useful discussions with Ravi Sheth and
especially Rein van de Weygaert whose detailed comments were
invaluable.



\appendix

\section{Estimate of Matter Flow in Void Formation}
We suggest the following simple estimate of the comoving distance
traveled by a fluid element in the process of the formation of a
spherical void of radius $R$.  We assume: (i) the initial density is
the mean density, (ii) the final density is uniform in the void and is
a fraction $f <1$ of the mean, (iii) the motion is purely radial, and
(iv) all the mass is accumulated on an infinitesimally thin shell of
radius $R$.

The model is purely kinematic and makes no assumptions regarding the
dynamics. It corresponds to the following mapping:
\begin{equation}
r=
\cases{
q+A\,q \qquad \mathrm{if}\quad q+Aq \le R, \quad (q \le q_c), \cr
R\qquad \qquad \; \mathrm{if}\quad q+Aq >R, \quad (q>q_c), \cr}
\end{equation}
where $q$ and $r$ are the initial and final positions of a fluid
element. The amplitude $A$ is determined by the value of $f$:
$A=f^{-1/3}-1$. The critical value $q_c$ where the mapping changes its
form is $q_c=R/(1+A)=f^{1/3}R$.

The mean distance traveled by a fluid element is 
\begin{eqnarray}
\bar{d}&=&{3 \over 4\pi R^3} 
\left[ \int\limits_0^{q_c} (Aq) 4\pi q^2  dq 
+\int\limits_{q_c}^R (R-q)4\pi q^2 dq \right] \\
&=&
{1-f \over 4}R,
\end{eqnarray}
which obviously understates the actual mean distance due to assumption
(iv).  The maximum distance traveled by fluid elements is
$d_{max}=(1-f^{1/3})R$.  If $f=0.1$ then $\bar{d} \approx 0.22$ and
$d_{max} \approx 0.54$.

Denoting the fraction of the distance by $\xi=d/R$ one can obtain
the pdf of the distances traveled by the fluid elements
\begin{eqnarray}
&&dF=3\left[1-2\xi+\left(1+{f \over (1-f^{1/3})^3}\right) \xi^2 \right]d\xi,\\
&&\qquad 0 \le \xi \le 1-f^{1/3}.\nonumber
\end{eqnarray}

\end{document}